\documentclass[12pt,a4paper,dvips]{article}
\usepackage{psfig,epsfig,wrapfig,times,mathptm} 
\renewcommand{\section}[1]{\vspace{6pt} \noindent\mbox{#1} \newline \noindent}
\renewcommand{\subsection}[1]{\vspace{6pt} \noindent\mbox{\underline{#1}} 
\newline \noindent}
\renewcommand{\subsubsection}[1]{\vspace{6pt} \noindent\mbox{\underline{#1}}
\noindent}

\newfont{\sansb}{cmssbx10}
\newfont{\sans}{cmss10}

\def \gray     {$\gamma$-ray }
\def \grays    {$\gamma$-rays }

\def \sig      {$\sigma$ }
\def \deg      {$^o$}

\begin{document}
{\center \LARGE COMPTEL OBSERVATIONS OF 3C~279 DURING THE FIRST 4 YEARS OF THE
 CGRO-MISSION \vspace{6pt}\\}
W. Collmar$^1$, J. J. Blom$^2$, K. Bennett$^4$, H. Bloemen$^2$, W. Hermsen$^2$, J. Ryan$^3$, V. Sch\"onfelder$^1$, J. G. Stacy$^3$, O. R. Williams$^4$
\vspace{6pt}\\
{\it $^1$Max-Planck-Institut f\"ur extraterrestrische Physik,
         P.O. Box 1603, 85740 Garching, F.R.G.\\
$^2$SRON-Utrecht, Sorbonnelaan 2, 3584 CA Utrecht, The Netherlands\\
$^3$University of New Hampshire, ISEOS, Durham NH 03824, USA\\
$^4$Astrophysics Division, ESA/ESTEC, NL-2200 AG Noordwijk, The Netherlands \vspace{-12pt}\\}
{\center ABSTRACT\\}
The COMPTEL experiment aboard the Compton Gamma-Ray Observatory (CGRO) has 
observed the gamma-ray blazar 3C 279 several times between April 1991 and 
September 1995. This paper reports on a consistent analysis of these
observations using the most recent COMPTEL data analysis tools. 
Detections and non-detections of 3C 279 along
the CGRO-mission indicate a time-variable MeV-flux. Spectral variability is
indicated as well, however can not be significantly proven by spectral
fitting. The average MeV-spectrum of 3C~279, as measured by COMPTEL over the 
four-year period, is consistent with 
a photon power-law slope of $\sim$ -1.9. This spectrum smoothly connects to the simultaneous 30~MeV to 10~GeV spectrum obtained from an analysis of
combined EGRET observations. No spectral break is 
required to fit the MeV- to GeV-spectrum of 3C~279.

\setlength{\parindent}{1cm}
\section{INTRODUCTION}
Shortly after the launch of the Compton Gamma-Ray Observatory (CGRO), 
the blazar-type quasar 3C~279 was detected as an emitter of 
\grays (mainly above 100~MeV) by the EGRET experiment aboard CGRO (Kniffen et al. 1993). This detection stimulated a search for 3C~279 in the 
contemporaneous COMPTEL data (for an instrument decription see Sch\"onfelder et al. 1993) between 0.75 and 30~MeV, which led to the discovery of 3C~279 at  MeV-energies as well (Hermsen et al. 1993, Williams et al. 1995). 3C~279 was redected by COMPTEL roughly two years later (Collmar et al. 1995).

Because only parts of the COMPTEL Virgo observations had been
analysed in detail,  
we have started to consistently analyse all of them with emphasis on 
3C~279 by applying 
the most recent COMPTEL analysis techniques and tools. The 
main objective is to study the MeV-properties of this blazar systematically,
and compare them to results in neighbouring energy bands (mainly EGRET). It is known that the MeV-band is an interesting region for 
blazar spectra (e.g. maximum energy release, spectral breaks) from which 
insights in the physical emission mechanisms might be derived.  
In this 
paper we shall report first results of these analyses covering the time
period between April '91 and September '95, which corresponds to 
the Phases I, II, III, and Cycle IV of the CGRO observations.

\section{OBSERVATIONS}
Between April '91 and September '95 COMPTEL was pointed towards
the Virgo region of the sky several times. All observations for which 3C~279
was within 25\deg of the pointing direction have been included in our
analysis resulting in a total (uncorrected) observation time of 136 days, 
which  converts to an effective observation time 
(including efficiency corrections and data losses) of $\sim$40 days. 
Table \ref{VPs} gives the 
relevant dates of the different CGRO viewing periods (VPs).  

\begin{table}[h]
\vspace{-12pt}
\caption{COMPTEL observations of the Virgo region up to the end of CGRO Cycle~IV. The fourth column gives the angular separation between the
 location of 3C~279
and the pointing direction. Subsequent VPs have been combined in the table.}\label{VPs}
\vspace{12pt}
\begin{center}
\begin{tabular}{ccccc}
\hline\hline
  VP  & Obs. Time            & Dur. & Separ.  & CGRO Phase\\
  \#  & yy/mm/dd - yy/mm/dd  & [days] &  $[ ^{\circ}]$ & \\
\hline
  3          & 91/06/15 - 91/06/28  & 13 & 8.8       & I    \\
 11          & 91/10/03 - 91/10/17  & 14 & 8.5       & I    \\
 204 - 206   & 92/12/22 - 93/01/12  & 21 & 7.2       & II   \\
 304 - 306   & 93/10/19 - 93/11/09  & 21 &  $\sim$14 & III  \\
 307 - 308.6 & 93/11/09 - 93/12/01  & 11 & $\sim$19.5 & III \\
 311 - 313   & 93/12/13 - 94/01/03  & 17 & $\sim$19  & III  \\
 405 - 408   & 94/11/29 - 95/01/10  & 32 & $\sim$7   & IV   \\
\hline
\hline
\end{tabular}
\end{center}
\end{table}

\section{DATA ANALYSIS}
We have applied the standard COMPTEL maximum-likelihood analysis method
(e.g. de Boer et al. 1992) to derive detection significances, fluxes, and 
flux errors of \gray sources in the four standard COMPTEL energy bands (0.75-1~MeV, 1-3~MeV, 3-10~MeV, 10-30~MeV), and a background modelling
technique 
which eliminates any source signature but preserves the general background 
structure (Bloemen et al. 1994). 
To derive source fluxes, we fitted 3C~279 simultaneously with 
further known \gray sources of the Virgo region (e.g. 3C~273)
in an iterative procedure.    
This iterative approach leads to a simultaneous determination of
the fluxes of several potential sources and a background model which takes
into account the presence of sources.

\section{RESULTS}
COMPTEL has significantly ($\geq$3\sig) detected 3C~279 at the beginning 
of its mission (Figure 1) mainly at energies above 3~MeV. The source was not detected during the 3 observational weeks on Virgo (VPs 204 - 206) in CGRO
Phase II, but was redetected in Phase III (Figure 1).
In contrast to Phase~I, a significant detection is derived in the 1-3~MeV band only.
Roughly one year later in CGRO Phase IV, 3C~279 was again invisible for 
COMPTEL during an observation of 4.5 weeks. 

\def\bbllx{ 2.0cm}
\def\bblly{ 8.5cm}
\def\bburx{19.5cm}
\def\bbury{25.9cm}
\def\height{8.0cm}
\def\bspace{-8.0cm}

\begin{figure} [bhtp]
\epsfig{figure=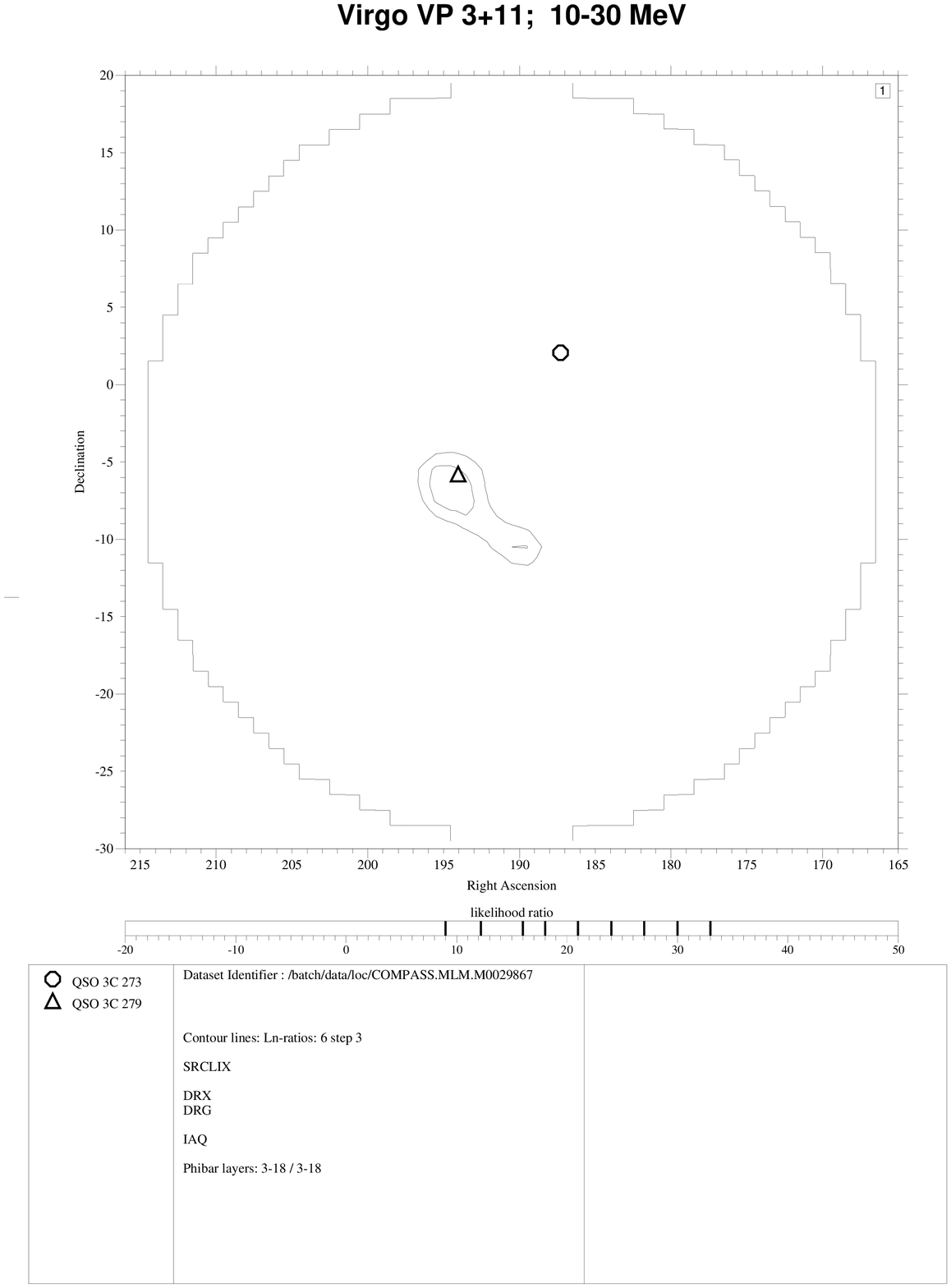,height=\height,bbllx=\bbllx,bblly=\bblly,bburx=\bburx,bbury=\bbury,clip=}

\vspace*{\bspace}\hfill
\psfig{figure=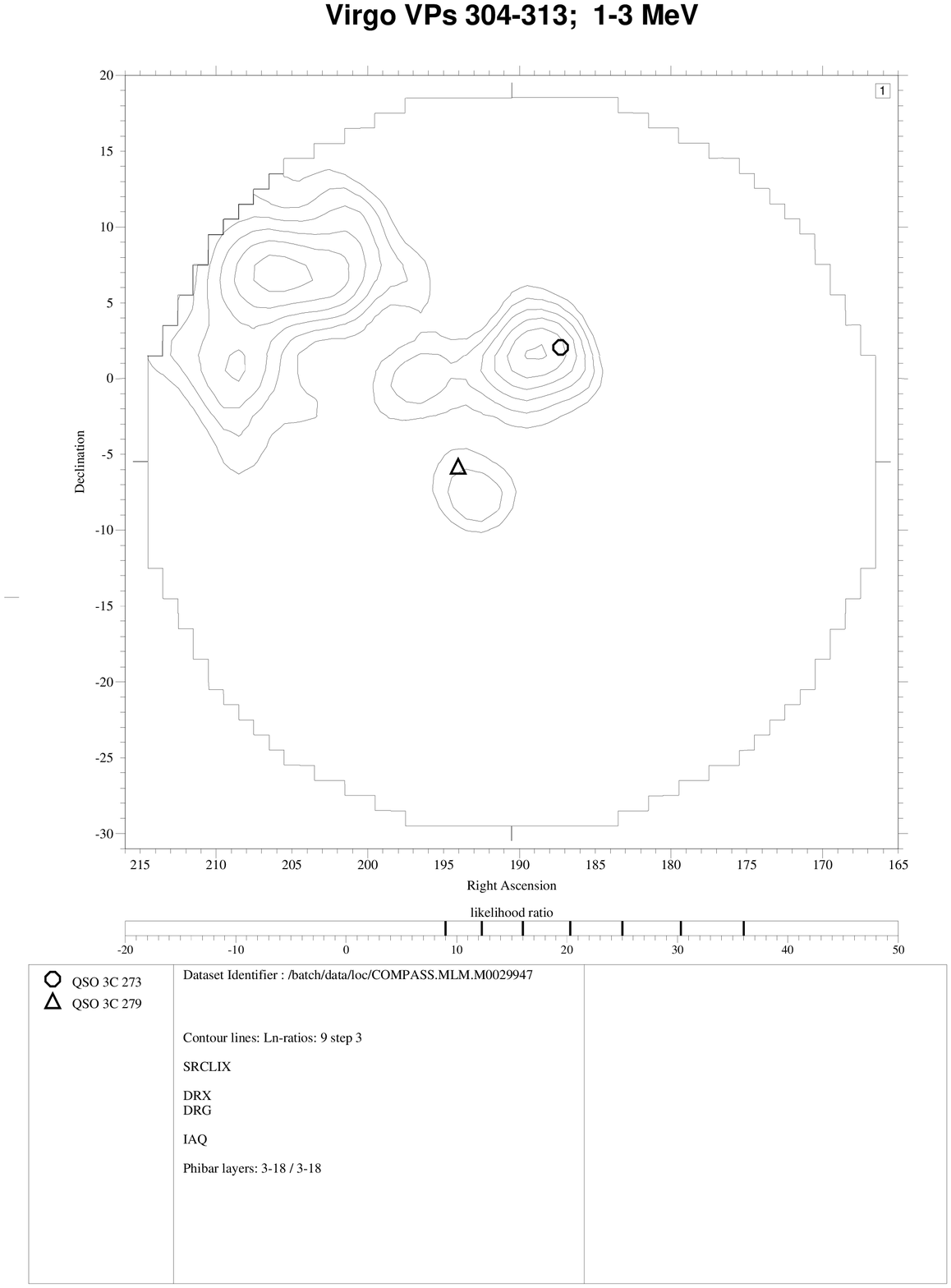,height=\height,bbllx=\bbllx,bblly=\bblly,bburx=\bburx,bbury=\bbury,clip=}

\vspace{-0.4cm}
\caption[]{\it
COMPTEL Phase I 10-30~MeV (left) and Phase III 1-3~MeV (right) Virgo maps (detection significances). The contour lines start at a detection
 significance of
3$\sigma$ ($\chi^{2}_{1}$-statistics) with a step of 0.5$\sigma$. The evidence for 3C~279 ($\triangle$) in both maps is visible. The location of the quasar 3C~273 (o) is indicated as well. In addition to 3C~279 and 3C~273 there is evidence for an unknown source in the 1-3~MeV map at $\alpha\sim$205\deg and $\delta\sim$+7\deg. Taking into account the number of trials, other features
of this map are not significant (Collmar et al. 1995).}
\end{figure}

\begin{wrapfigure} [23] {r} {9.5cm}
\vspace{-1cm}
\epsfig{figure=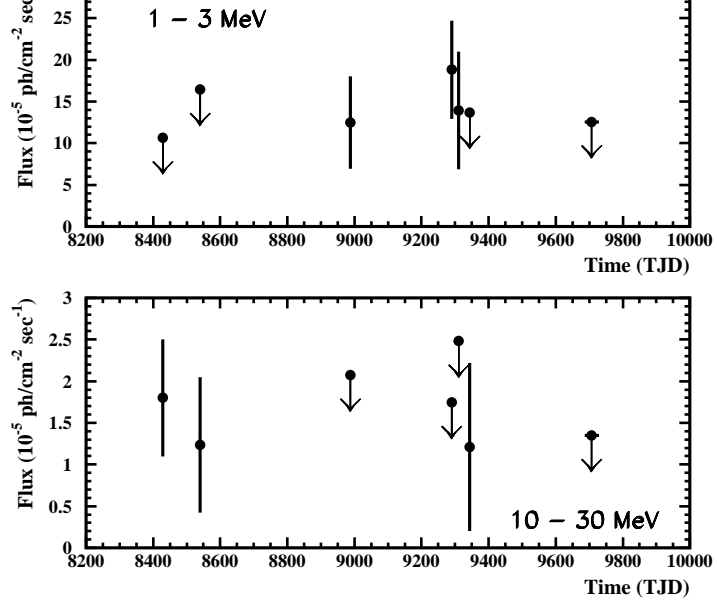,width=9.5cm,height=9.5cm,angle=0}
\vspace{-1.0cm}
\caption[]{\it Time history of 3C~279 as measured by COMPTEL in
the 1-3~MeV and 10-30~MeV
bands. The data points correspond to the observational periods given in Table \ref{VPs}. The error bars are 1$\sigma$ and the upper limits are 2$\sigma$.
An upper limit is drawn when the significance of an individual flux value is below 1$\sigma$.}  
\end{wrapfigure}

This behaviour is illustrated by the long-term lightcurves of the 
quasar. The fluxes and upper limits of the COMPTEL 1-3~MeV and 10-30~MeV bands for different observational periods are shown in Figure~2. In the 10-30~MeV
band positive flux values are mainly observed during Phase I, 
in the 1-3~MeV band however, mainly during CGRO Phase~III.

In order to investigate the energy spectra of 3C~279 we have generated
deconvolved, background-subtracted source fluxes in the four standard
COMPTEL energy bands by application of the method
described above. These spectra were generated for the observational
periods for which 3C~279 was detected, the combined Phase I observations (VPs 3+11)and the combined Phase III observations, and for the sum of all 
data, to derive a four-year average MeV-spectrum of 3C~279. 
To determine the spectral shape, we fitted a simple
power-law function of the form 
\begin{equation}
\label{SPL}
I(E) = I_{0}  (E/E_{0})^{-\alpha} \,\, {\rm photons\,\,cm}^{-2} {\rm s}^{-1} {\rm MeV} ^{-1}
\end{equation}  
with the parameters $\alpha$ (photon spectral index) and $I_{0}$ (intensity at the normalisation energy $E_{0}$). $E_{0}$ was chosen that the two free parameters are minimally correlated. 
The 1$\sigma$-errors on the parameters are derived for two parameters of 
interest by adding 2.3 to the
minimum $\chi^{2}$-value (Lampton et al. 1976).
The COMPTEL spectra together with the best-fit model are shown in Figure~3, 
and the fit results are given in Table~2.

\begin{table}[bth]
\vspace{-12pt}
\caption[]{Results of the power-law fits of the different spectra.
The errors on the fitparameters are 1$\sigma$. The last line gives the fit results of the combined COMPTEL/EGRET spectrum.}\label{FITS}
\vspace{12pt}
\begin{center}\begin{tabular}{cccccc}
\hline\hline
  Obs & Range & E$_0$ & I$_{0}$(E$_0$)   &  $\alpha$ & $\chi^{2}_{red}$ \\
      & (MeV) & (MeV) & (ph cm$^{-2}$ s$^{-1}$ MeV$^{-1}$) & &   \\
\hline
Phase I    & 0.75-30 & 6.0 & (2.9$\pm^{2.0}_{2.4}$) 10$^{-6}$ & 1.2$\pm^{0.7}_{0.9}$ & 1.6 \\
Phase III  & 0.75-30 & 3.0 & (12.9$\pm^{7.1}_{7.6}$) 10$^{-6}$  & 2.2$\pm^{1.0}_{0.6}$ & 0.1  \\
Phase I-IV & 0.75-30 & 3.0 & (11.6$\pm^{4.2}_{4.5}$) 10$^{-6}$ & 1.9$\pm^{0.4}_{0.4}$ & 1.1 \\    Phase I-IV & 0.75-10000&200& (2.15$\pm^{0.10}_{0.10}$) 10$^{-9}$& 2.04$\pm^{0.04}_{0.04}$& 1.5 \\
\hline\hline
\end{tabular}\end{center}\end{table}

The images (detections) indicate during Phase~I a "harder" spectrum than 
during Phase~III. This trend is supported by spectral fits as well. 
However, the spectral slopes are consistent within error bars. 
The average MeV-spectrum, measured over the four-year 
period, shows a spectral slope of roughly -1.9.
To check for a possible spectral break, we enlarged the energy range by 
generating an average four-year EGRET spectrum of 3C~279 contemporaneous to the COMPTEL one.  
The COMPTEL points smoothly connect to the EGRET spectrum (Figure 3).
No spectral break is obvious. The average
MeV- to GeV-spectrum of 3C~279 is consistent with a simple power-law
with a slope of $\sim$-2 (Table~2). 
By comparing non-simultaneous OSSE and EGRET data, McNaron-Brown et
al. (1995) also found the spectrum of 3C~279 to be consistent with a simple power-law shape with a photon index of $\sim$-1.9.

\begin{figure} [t]
\psfig{figure=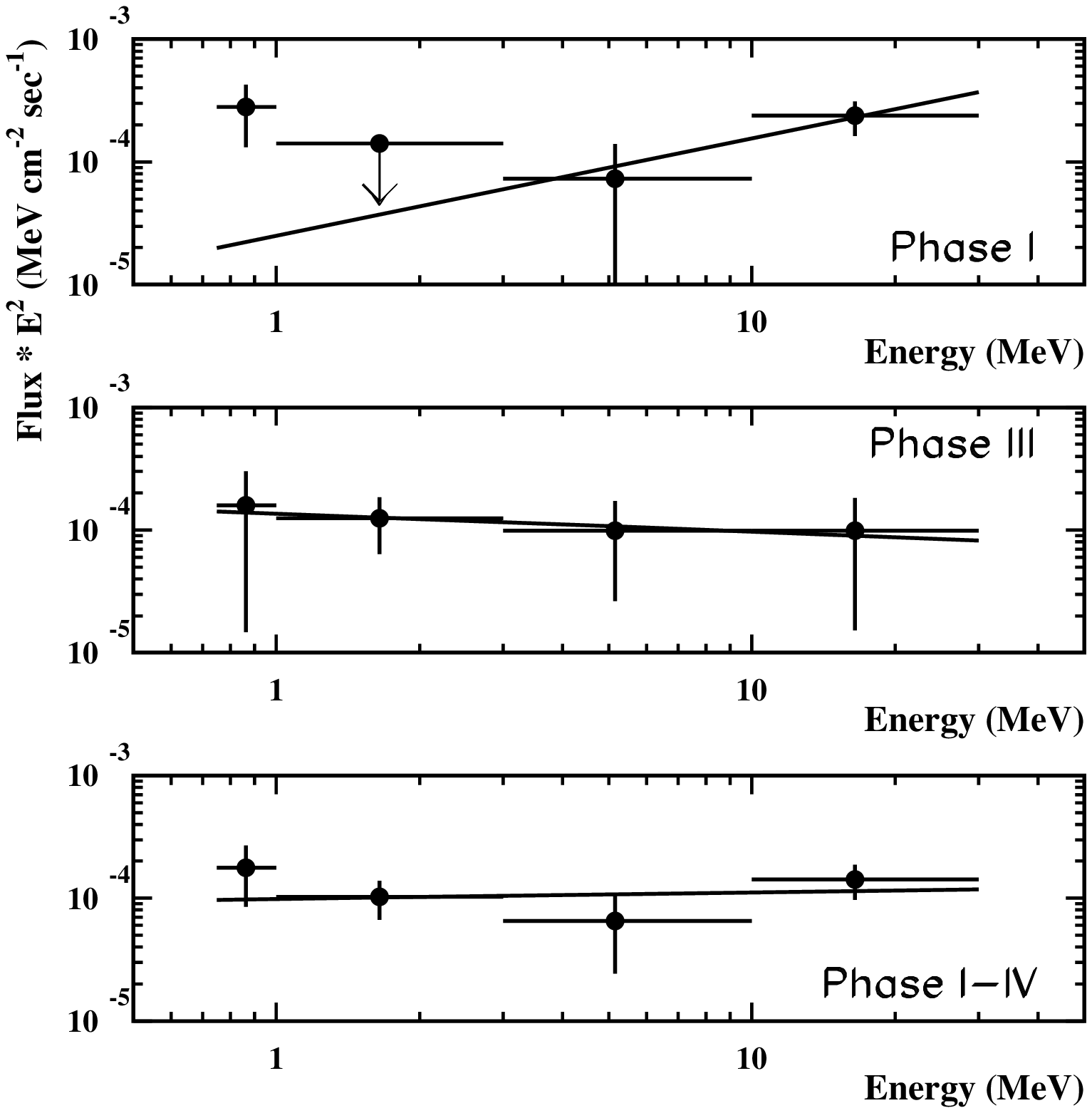,width=8.5cm,height=8.0cm,angle=0,clip=}

\vspace*{-8.0cm}\hfill
\psfig{figure=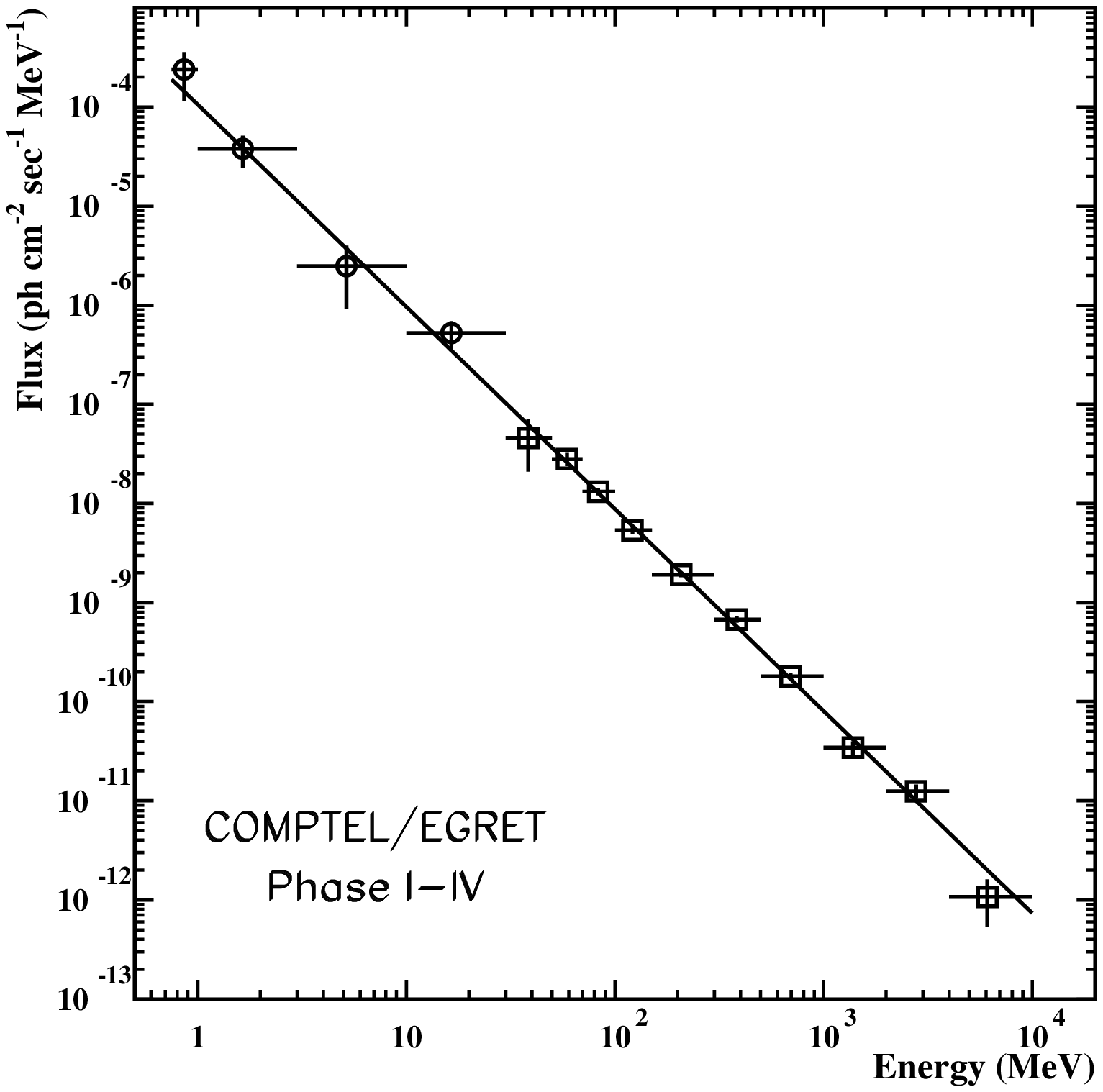,width=8.5cm,height=8.0cm,angle=0,clip=}
\vspace{-0.3cm}
\caption{\it Left: 3C~279 COMPTEL spectra for combined Phase I (VPs 3+11), 
combined Phase III, and the sum of all data. The spectra are shown as 
differential flux $\times$ E$^2$. The solid lines represent the best-fit 
power-law models. The error bars are 1\sig and the upper limits are 2\sig.
The fits have been performed by using the measured flux values and not
the upper limits shown. 
An upper limit is drawn when the significance of a flux value is below 1\sig.
Right: The contemporaneous EGRET/COMPTEL differential spectrum for the sum of all data together with the best-fit power-law model.
The error bars are 1\sig.}
\end{figure}

\section{SUMMARY}
We have started to consistently analyse the data of COMPTEL 3C~279 
observations during several years. Detections and non-detections of
3C 279 along the CGRO-mission indicate a time-variable MeV-flux.
Spectral variability is indicated as well, however can not be significantly proven by spectral fits. The average 3C~279 MeV-spectrum over four years 
is consistent with 
a photon power-law slope of $\sim$ -1.9, which smoothly connects to the simultaneous EGRET spectrum. No spectral break is 
required to fit the combined MeV- to GeV-spectrum of 3C~279.

\section{ACKNOWLEDGEMENTS}
This research was supported by the Deutsche Agentur f\"ur Raumfahrtangelegenheiten (DARA) under the grant 50 QV 90968, by NASA under   contract NASA-26645, and by the Netherlands Organisation for Scientific Research.

\section{REFERENCES}
\setlength{\parindent}{-5mm}
\begin{list}{}{\topsep 0pt \partopsep 0pt \itemsep 0pt \leftmargin 5mm
\parsep 0pt \itemindent -5mm}
\vspace{-15pt}
\item Bloemen H., Hermsen W., Swanenburg B.N. et al., ApJ Suppl. 92, 419 (1994). \item de Boer H., Bennett K., Bloemen H. et al., In: {\it Data Analysis in Astronomy IV}, eds. V. Di Gesu et al. (New York: plenum Press), 241 (1992).
\item Collmar W., Bennett K., Bloemen H. et al., MPE Report 261, 81 (1995).
\item Hermsen, W., Aarts, H.J.M., Bennett, K. et al., A\&A Suppl. 97, 97 (1993).
\item Kniffen, D. A., Bertsch D. L., Fichtel, C. E. et al., ApJ, 411, 133 (1993).
\item Lampton, M., Margon, B., Bowyer, S., ApJ 208, 177 (1976).
\item McNaron-Brown, K., Johnson, W. N., Jung, G. V. et al., ApJ 451, 575 (1995).
\item Sch\"onfelder, V., Aarts, H., Bennett, K. et al., ApJ Suppl. 86, 657 (1993).  
\item Williams O. R., Bennett K., Bloemen H. et al., A\&A 298, 33 (1995).  
\end{list}

\end{document}